\documentclass[epj,twocolumn]{webofc}
\usepackage[varg]{txfonts}   
\usepackage{graphicx}
\usepackage{url}
\usepackage{natbib}
\usepackage{subfigure}

\wocname{epj}
\woctitle{Seismology of the Sun and the Distant Stars 2016}
\begin{document}
\title{On the detectability of solar-like oscillations with the NASA \textit{TESS} mission}
%

\author{\firstname{Tiago L.} \lastname{Campante}\inst{1,2}\fnsep\thanks{\email{campante@bison.ph.bham.ac.uk}}}

\institute{
School of Physics and Astronomy, University of Birmingham, Edgbaston, Birmingham B15 2TT, UK
\and
Stellar Astrophysics Centre (SAC), Department of Physics and Astronomy, Aarhus University, Ny Munkegade 120,\\ DK-8000 Aarhus C, Denmark
}

\abstract{
The upcoming NASA \textit{TESS} mission will perform an all-sky survey for planets transiting bright nearby stars. In addition, its excellent photometric precision will enable asteroseismology of solar-type and red-giant stars. We apply a newly developed detection test along a sequence of stellar evolutionary tracks in order to predict the detectability of solar-like oscillations with \textit{TESS}.
}
\maketitle
%
\section{Introduction}
\label{sec:intro}
The \textit{Transiting Exoplanet Survey Satellite}\footnote{\url{https://tess.gsfc.nasa.gov}} \cite[\textit{TESS};][]{TESS1} is a NASA-sponsored Astrophysics Explorer mission that will perform an all-sky survey for planets transiting bright nearby stars. Its launch is currently scheduled for December 2017, with science operations due to start in March 2018. During the 2-year primary mission, \textit{TESS} will monitor the brightness of several hundred thousand main-sequence, low-mass stars over intervals ranging from one month to one year, depending mainly on a star's ecliptic latitude. Monitoring of these pre-selected target stars will be made at a cadence of {2\:{\rm min}}. Additionally, full-frame images (FFIs) will be recorded every {30\:{\rm min}}, which will allow a wider range of stars to be searched for transits. Being 10--100 times brighter than \textit{Kepler} targets and distributed over a solid angle that is nearly 300 times larger, \textit{TESS} host stars will be well suited for follow-up radial-velocity/spectroscopic observations from the ground. \textit{TESS} is expected to detect approximately 1700 transiting planets from $\sim\!2\!\times\!10^5$ pre-selected target stars \cite{TESS2}. Analysis of the FFIs will lead to the additional detection of several thousand planets orbiting stars that are not among the pre-selected targets.

Furthermore, \textit{TESS}'s excellent photometric precision, combined with its fine time sampling and long intervals of uninterrupted observations, will enable asteroseismology of solar-type and red-giant stars, whose dominant oscillation periods range from minutes to hours. In Section \ref{sec:detectability}, we apply a newly developed detection test along a sequence of stellar evolutionary tracks in order to predict the detectability of solar-like oscillations with \textit{TESS} across the Hertzsprung--Russell (H--R) diagram.

\section{Detectability of solar-like oscillations across the H--R diagram}
\label{sec:detectability}
New images will be acquired by each of the four \textit{TESS} cameras every 2 seconds. However, due to limitations in onboard data storage and telemetry, these 2-sec images will be stacked (before being downlinked to Earth) to produce two primary data products with longer effective exposure times: (i) subarrays of pixels centered on several hundred thousand pre-selected target stars will be stacked at a 2-min cadence, while (ii) FFIs will be stacked every {30\:{\rm min}}. 

Up to $20{,}000$ 2-min-cadence slots (or the equivalent to $\sim\!10\,\%$ of the pre-selected target stars) will be allocated to the \textit{TESS} Asteroseismic Science Consortium (TASC) over the course of the mission. In addition, a number of slots (notionally 1500) with faster-than-standard sampling, i.e., $20\:{\rm sec}$, will be reserved for the investigation of asteroseismic targets of special interest (mainly compact pulsators and main-sequence, low-mass stars).

In \cite{Campante16} we developed a simple test to estimate the detectability of solar-like oscillations in \textit{TESS} photometry of any given star. This test is in turn based on the test previously developed by \cite{Chaplin11} to estimate the detectability of solar-like oscillations in any given \textit{Kepler} target, which looked for signatures of the bell-shaped power excess due to the oscillations. In what follows, we apply the newly developed detection test along a sequence of stellar evolutionary tracks in order to predict the detectability of solar-like oscillations.

\begin{figure*}[!t]
\centering
  \subfigure[$T=27\:{\rm d}$, $\sigma_{\rm sys}\!=\!0\:{\rm ppm\,hr^{1/2}}$.]{%
  \includegraphics[width=.5\textwidth]{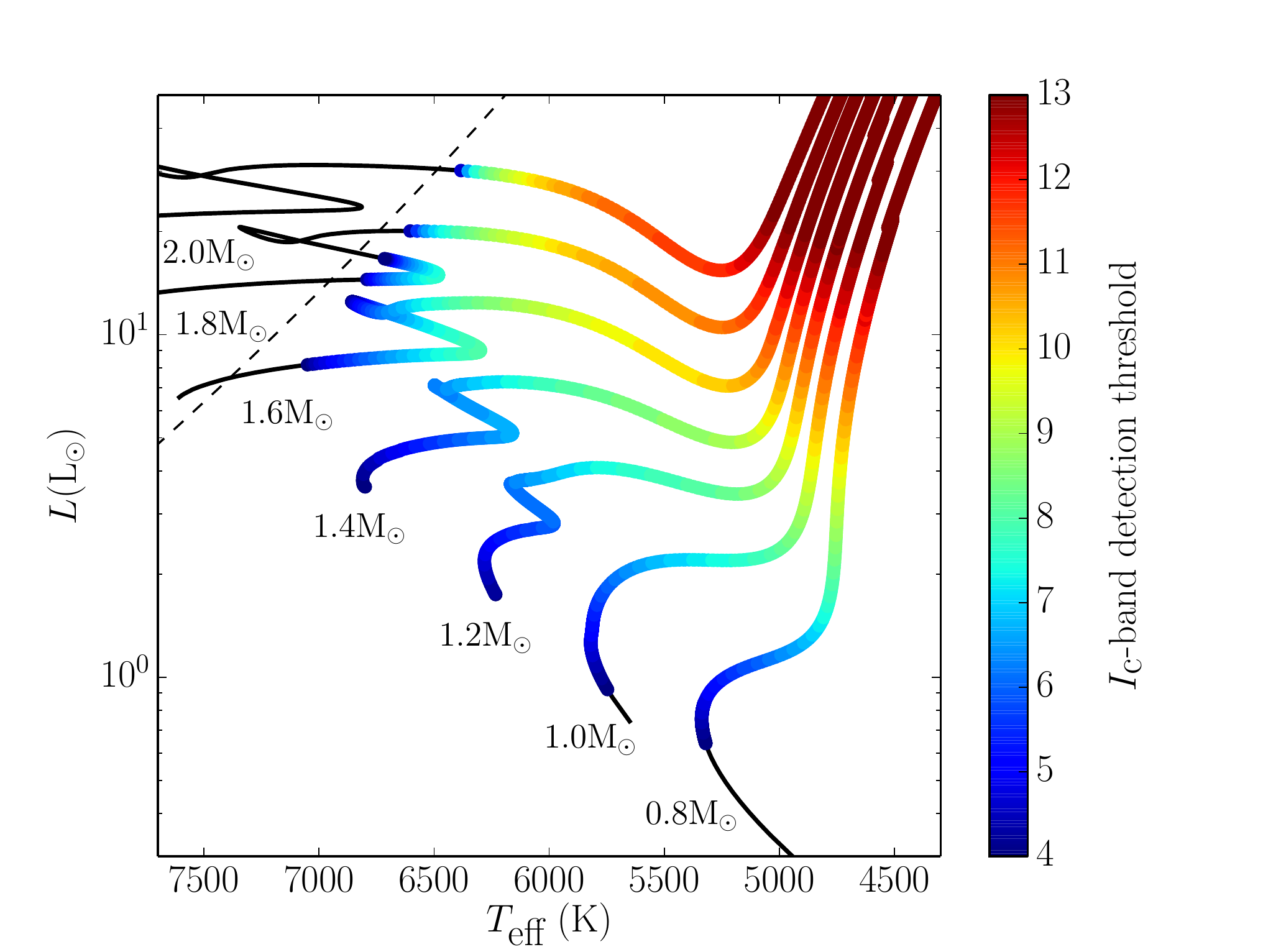}}\hfill
  \subfigure[$T=27\:{\rm d}$, $\sigma_{\rm sys}\!=\!60\:{\rm ppm\,hr^{1/2}}$.]{%
  \includegraphics[width=.5\textwidth]{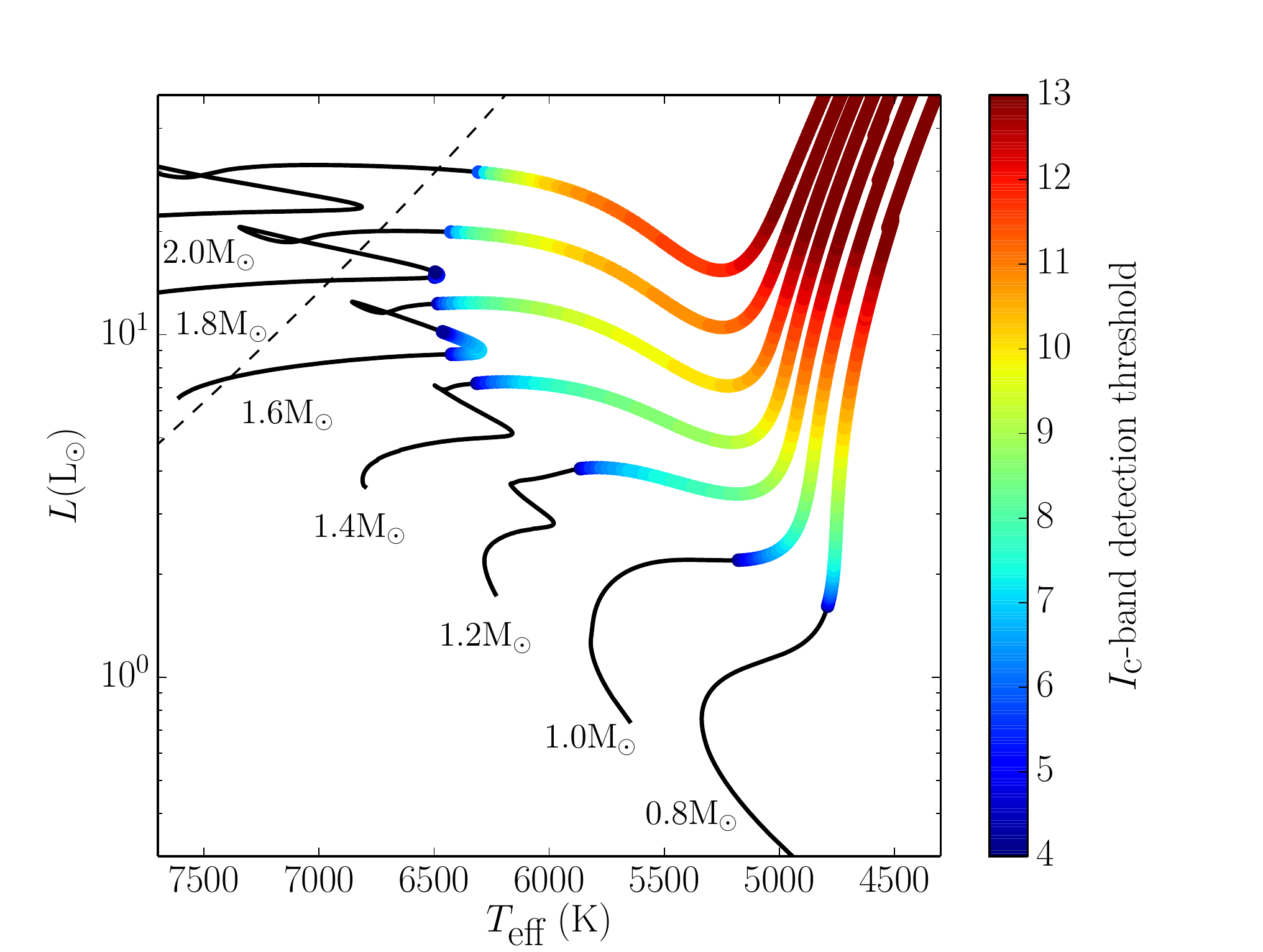}}\\
  \subfigure[$T=351\:{\rm d}$, $\sigma_{\rm sys}\!=\!0\:{\rm ppm\,hr^{1/2}}$.]{%
  \includegraphics[width=.5\textwidth]{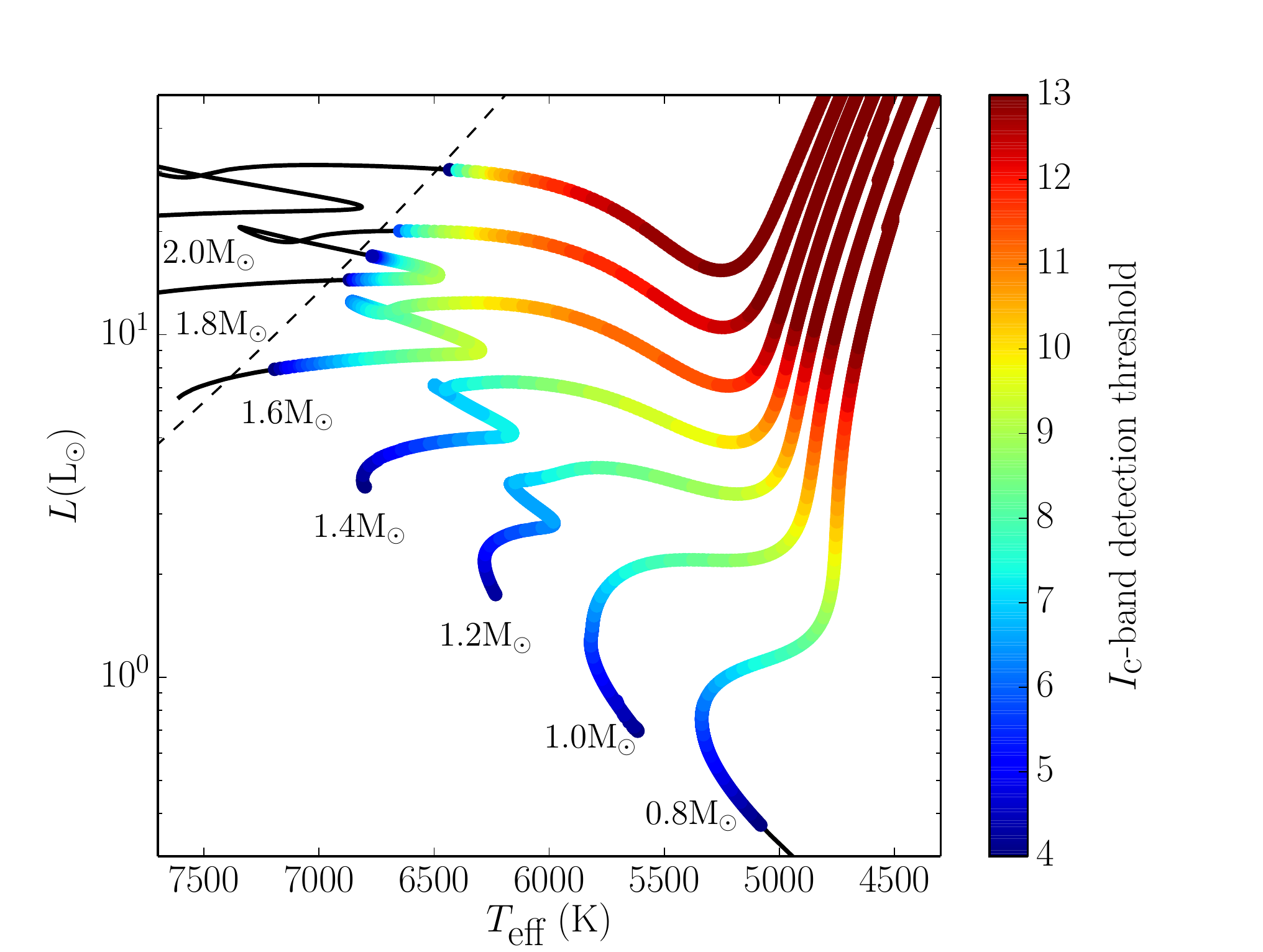}}\hfill
  \subfigure[$T=351\:{\rm d}$, $\sigma_{\rm sys}\!=\!60\:{\rm ppm\,hr^{1/2}}$.]{%
  \includegraphics[width=.5\textwidth]{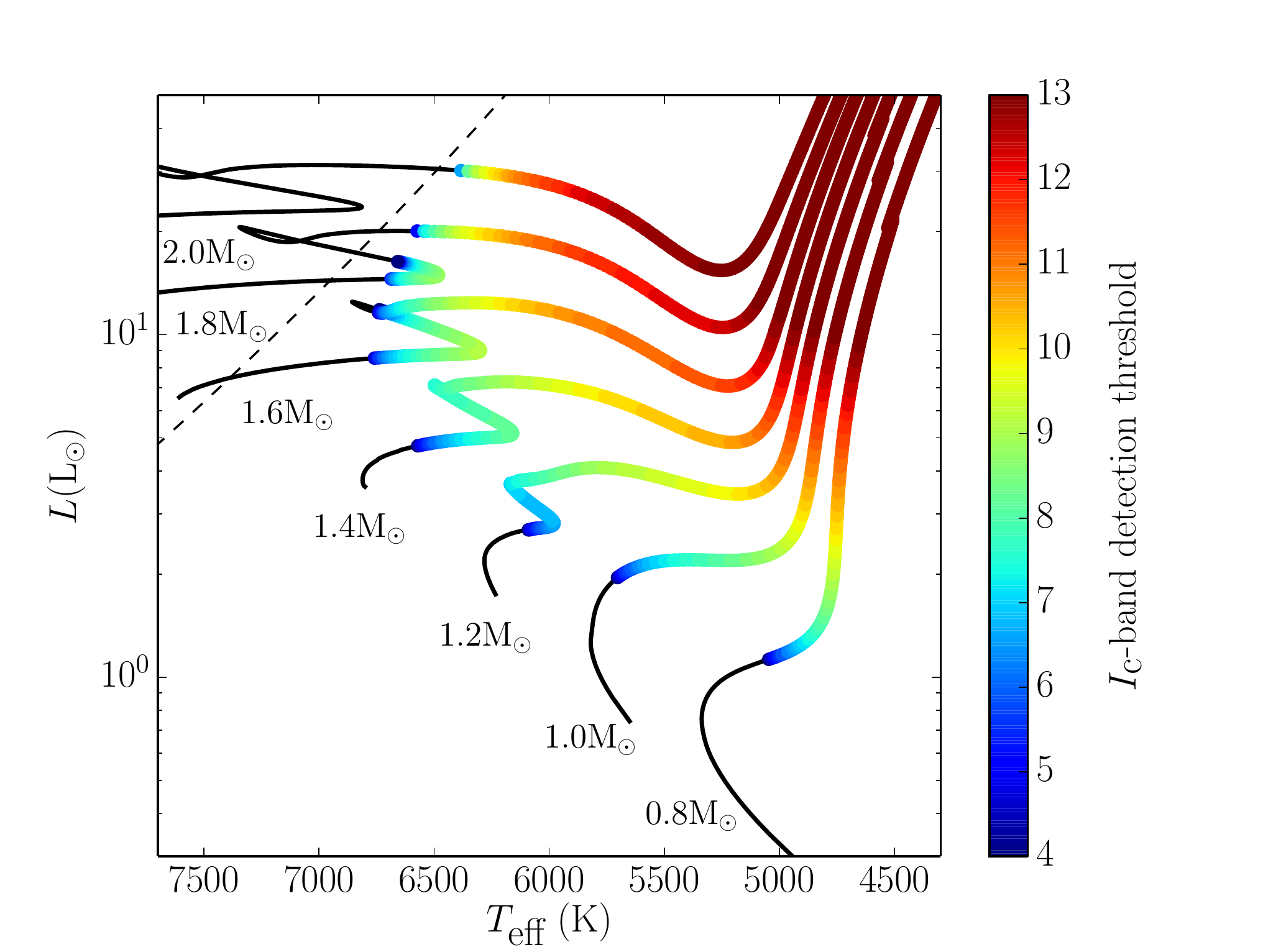}}
  \caption{\small Detectability of solar-like oscillations with \textit{TESS} across the H--R diagram for a cadence of $\Delta t\!=\!2\:{\rm min}$. $I_{\rm C}$-band detection thresholds are color-coded (no detection is possible along those portions of the tracks shown as a thin black line). The slanted dashed line represents the red edge of the $\delta$ Scuti instability strip. The several panels consider different combinations of the length of the observations ($T$) and systematic noise level ($\sigma_{\rm sys}$), as indicated. This figure was originally published in \cite{Campante16}.}\label{fig:stellar_tracks_120s_lum}
\end{figure*}

\begin{figure*}[!t]
\centering
  \subfigure[$T=27\:{\rm d}$, $\sigma_{\rm sys}\!=\!60\:{\rm ppm\,hr^{1/2}}$.]{%
  \includegraphics[width=.5\textwidth]{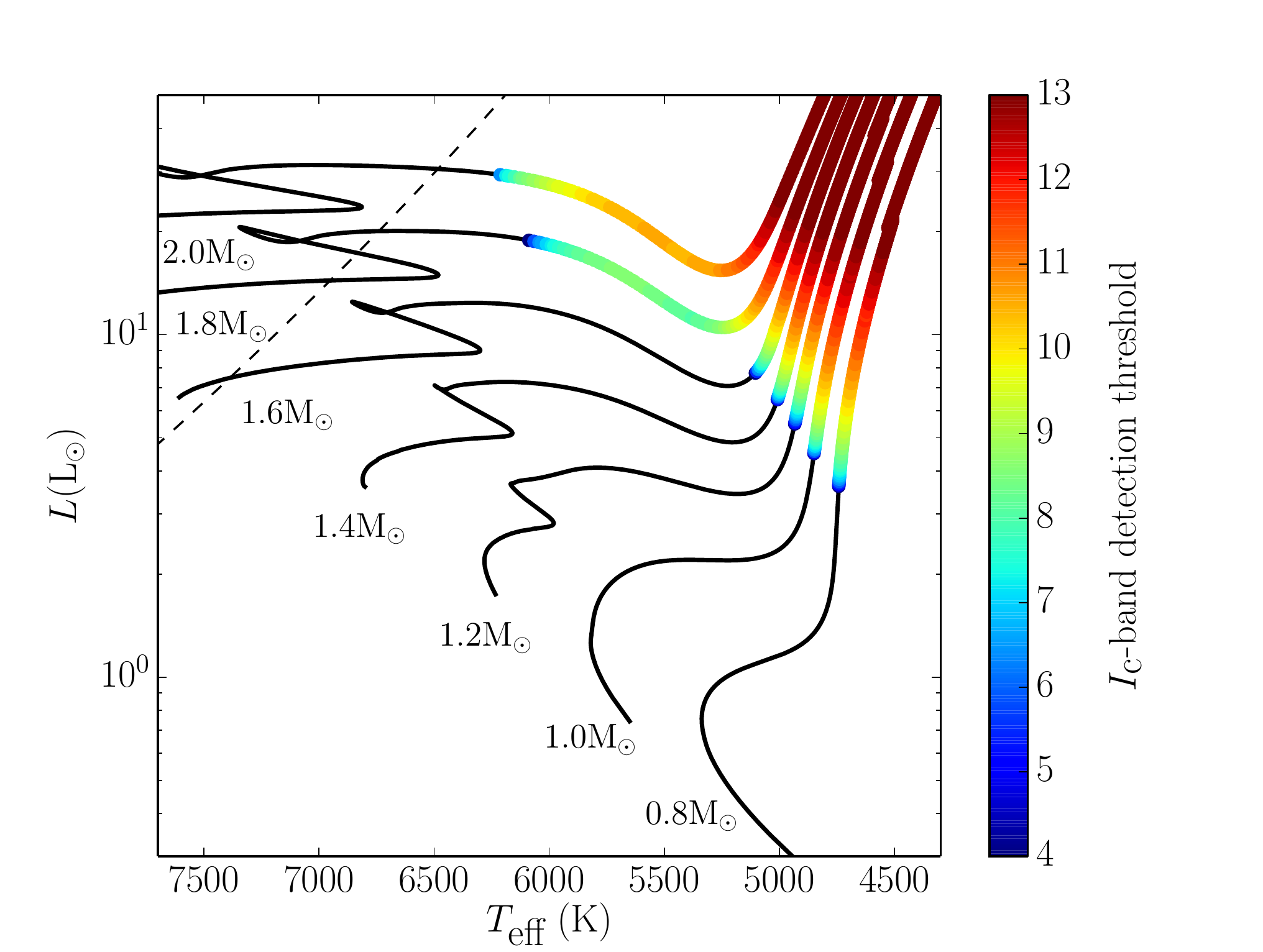}}\hfill
  \subfigure[$T=351\:{\rm d}$, $\sigma_{\rm sys}\!=\!60\:{\rm ppm\,hr^{1/2}}$.]{%
  \includegraphics[width=.5\textwidth]{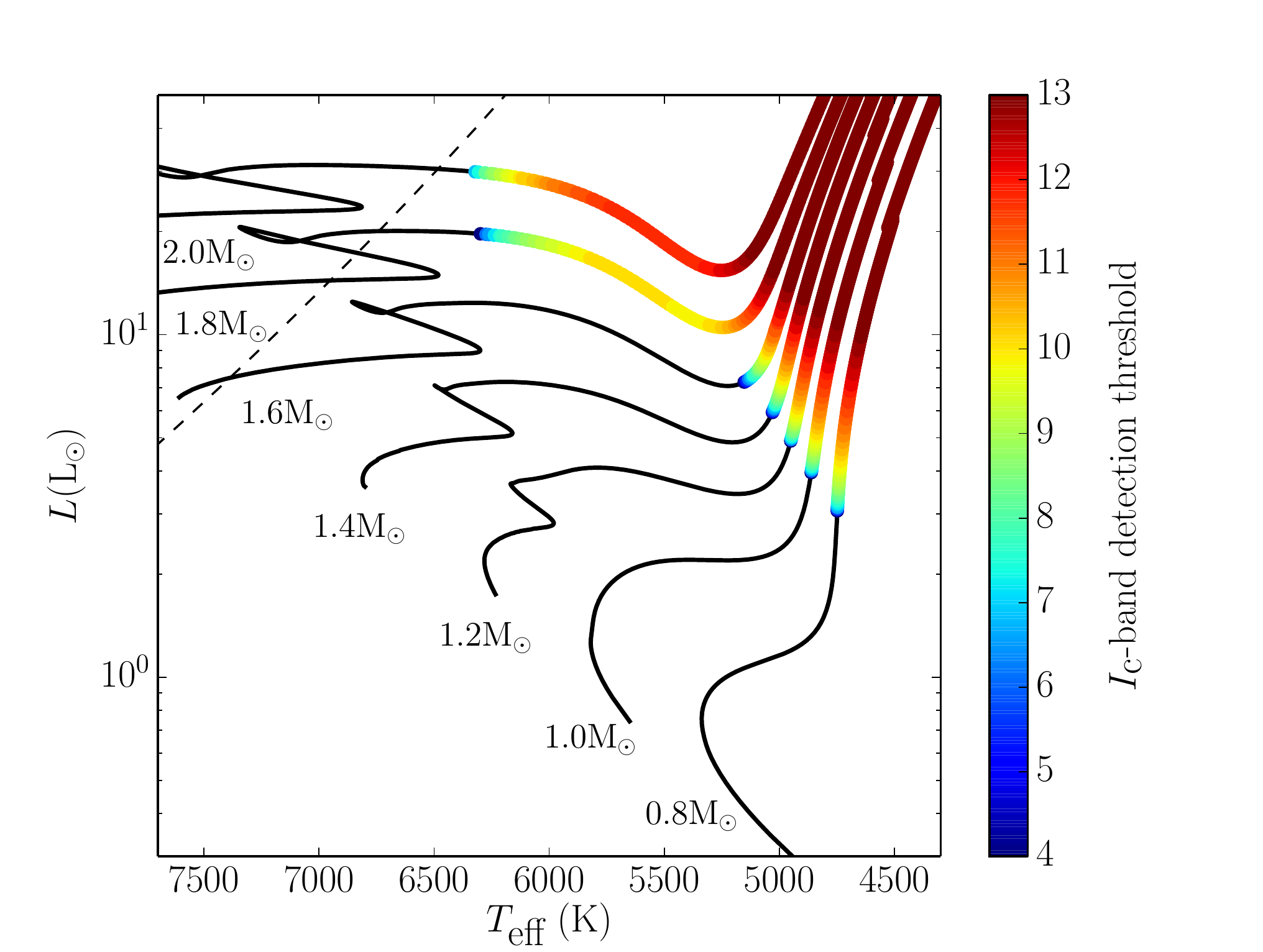}}
  \caption{\small Detectability of solar-like oscillations with \textit{TESS} across the H--R diagram for a cadence of $\Delta t\!=\!30\:{\rm min}$. $I_{\rm C}$-band detection thresholds are color-coded (no detection is possible along those portions of the tracks shown as a thin black line). The slanted dashed line represents the red edge of the $\delta$ Scuti instability strip. The two panels consider different lengths of the observations ($T$) and a systematic noise level of $\sigma_{\rm sys}\!=\!60\:{\rm ppm\,hr^{1/2}}$, as indicated. This figure was originally published in \cite{Campante16}.}\label{fig:stellar_tracks_1800s_lum}
\end{figure*}

\begin{figure*}[!t]
\centering
  \subfigure[$T=27\:{\rm d}$, $\sigma_{\rm sys}\!=\!60\:{\rm ppm\,hr^{1/2}}$.]{%
  \includegraphics[width=.5\textwidth]{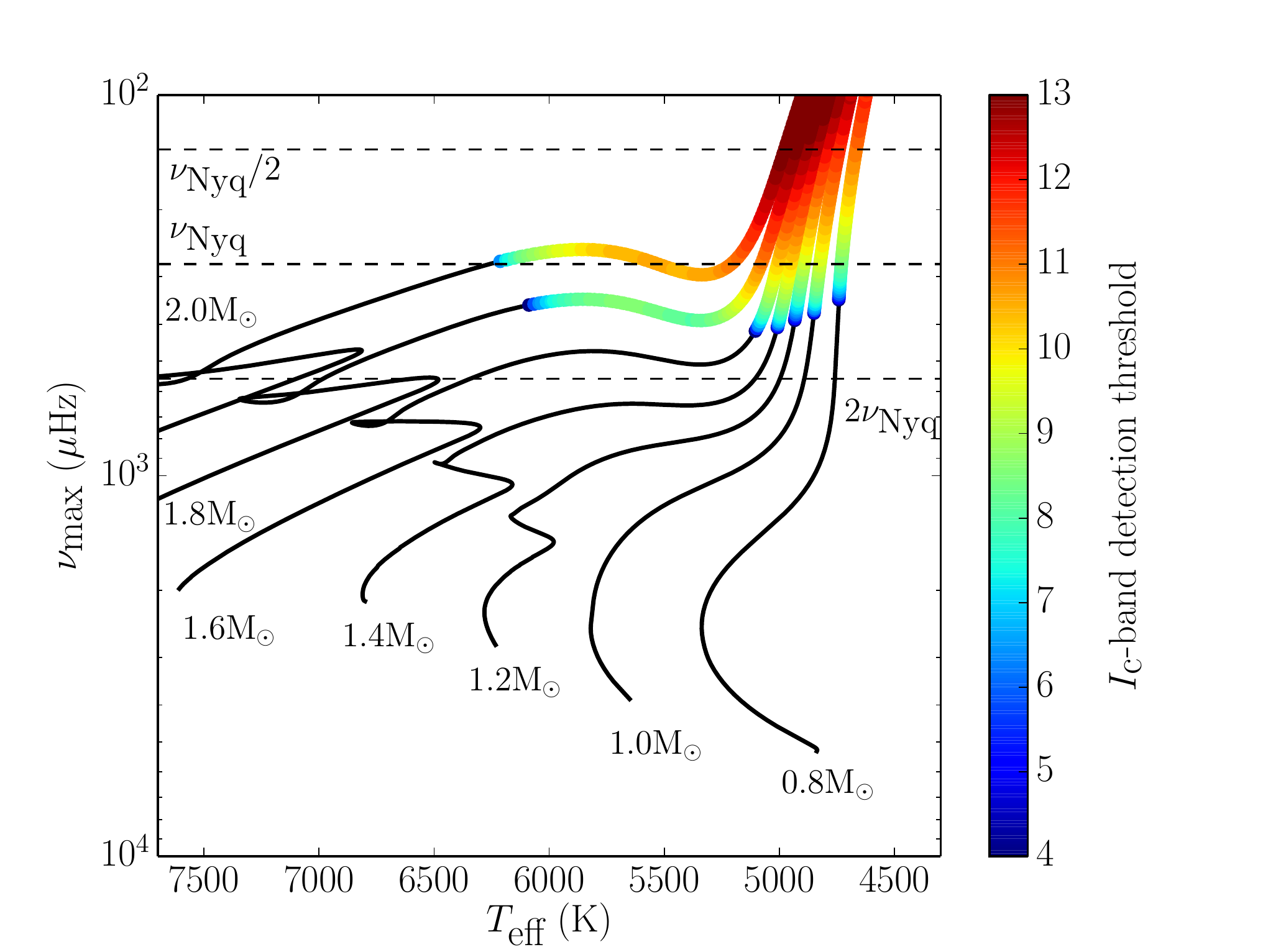}}\hfill
  \subfigure[$T=351\:{\rm d}$, $\sigma_{\rm sys}\!=\!60\:{\rm ppm\,hr^{1/2}}$.]{%
  \includegraphics[width=.5\textwidth]{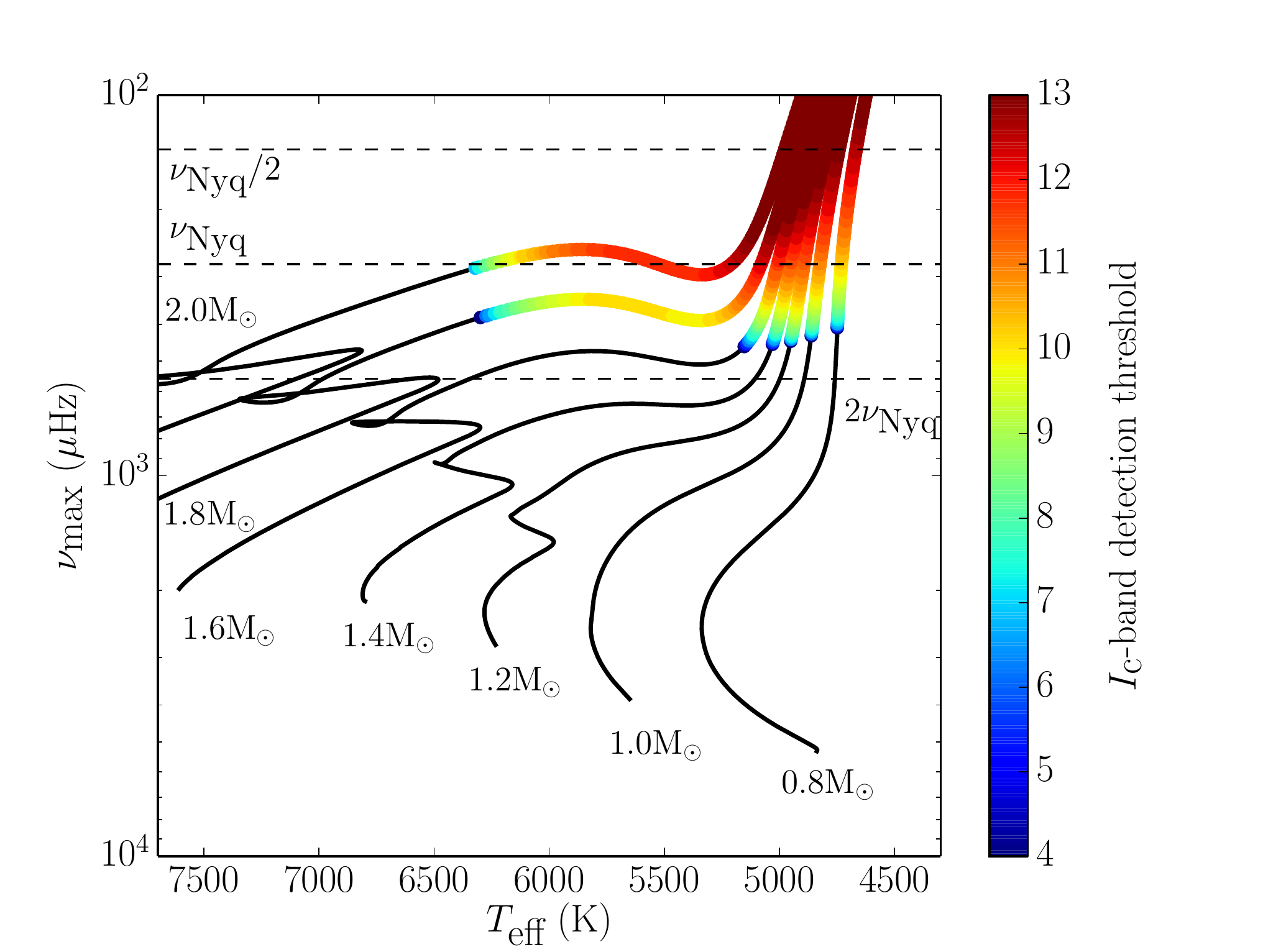}}
  \caption{\small Detectability of solar-like oscillations with \textit{TESS} across an asteroseismic H--R diagram for a cadence of $\Delta t\!=\!30\:{\rm min}$. Note that the frequency of maximum oscillation amplitude, $\nu_{\rm max}$, is now plotted along the vertical axis and not luminosity. Horizontal dashed lines indicate $\nu_{\rm Nyq}/2$, $\nu_{\rm Nyq}$ and $2\nu_{\rm Nyq}$, where $\nu_{\rm Nyq}$ is the Nyquist frequency. $I_{\rm C}$-band detection thresholds are color-coded (no detection is possible along those portions of the tracks shown as a thin black line). The two panels consider different lengths of the observations ($T$) and a systematic noise level of $\sigma_{\rm sys}\!=\!60\:{\rm ppm\,hr^{1/2}}$, as indicated. This figure was originally published in \cite{Campante16}.}\label{fig:stellar_tracks_1800s_numax}
\end{figure*}

Figures \ref{fig:stellar_tracks_120s_lum}--\ref{fig:stellar_tracks_1800s_numax} illustrate the detectability of solar-like oscillations with \textit{TESS}. We focus on that portion of the H--R diagram populated by solar-type and (low-luminosity) red-giant stars, bound at high effective temperatures by the red edge of the $\delta$ Scuti instability strip. The detection code was applied along a sequence of solar-calibrated stellar evolutionary tracks spanning the mass range 0.8--$2.0\,{\rm M}_\odot$ (in steps of $0.2\,{\rm M}_\odot$).

In Figure \ref{fig:stellar_tracks_120s_lum} we consider two different observing lengths and a cadence of $\Delta t\!=\!2\:{\rm min}$. Further assuming a systematic noise level\footnote{This is an engineering requirement imposed on the design of the \textit{TESS} photometer and not an estimate of the anticipated systematic noise level on 1-hour timescales. In the actual data it is expected that this systematic term will be rather small in the frequency range occupied by the oscillations. Herein, we explore the implications of having $\sigma_{\rm sys}\!=\!0\:{\rm ppm\,hr^{1/2}}$ (ideal case) and $\sigma_{\rm sys}\!=\!60\:{\rm ppm\,hr^{1/2}}$ (regarded as a worst-case scenario).} of $\sigma_{\rm sys}\!=\!60\:{\rm ppm\,hr^{1/2}}$, detection of solar-like oscillations in main-sequence stars will not be possible for $T\!=\!27\:{\rm d}$. Increasing the observing length to $T\!=\!351\:{\rm d}$ (relevant for stars near the ecliptic poles) may lead to the marginal detection of oscillations in (very bright) main-sequence stars more massive than the Sun. In both cases, detection of oscillations in subgiant and red-giant stars is nonetheless made possible, owing to their higher intrinsic amplitudes. As one would expect, this situation is significantly improved as the systematic noise level is brought down to $\sigma_{\rm sys}\!=\!0\:{\rm ppm\,hr^{1/2}}$, with detections now being made possible for the brightest main-sequence stars over a range of masses.

The longer 30-min cadence is considered in Figures \ref{fig:stellar_tracks_1800s_lum} and \ref{fig:stellar_tracks_1800s_numax}, where we have assumed a systematic noise level of $\sigma_{\rm sys}\!=\!60\:{\rm ppm\,hr^{1/2}}$. FFIs will allow detecting oscillations in red-giant stars down to relatively faint magnitudes. Furthermore, it becomes apparent from Figure \ref{fig:stellar_tracks_1800s_numax} that it should be possible to detect oscillations in the super-Nyquist regime for the brightest red giants. 

\begin{figure*}[!t]
\centering
  \subfigure[Kepler-56 as observed by \textit{Kepler}.]{%
  \includegraphics[width=.5\textwidth]{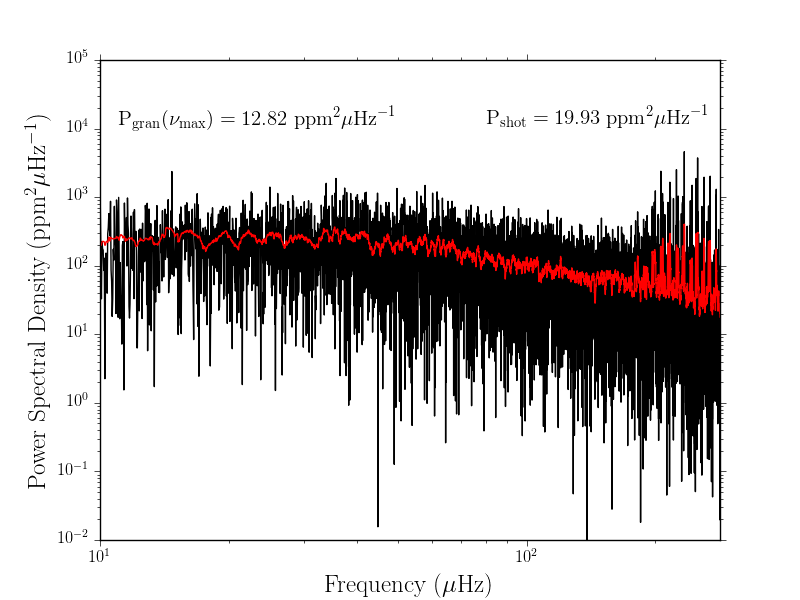}}\hfill
  \subfigure[Kepler-56 as observed by \textit{TESS}.]{%
  \includegraphics[width=.5\textwidth]{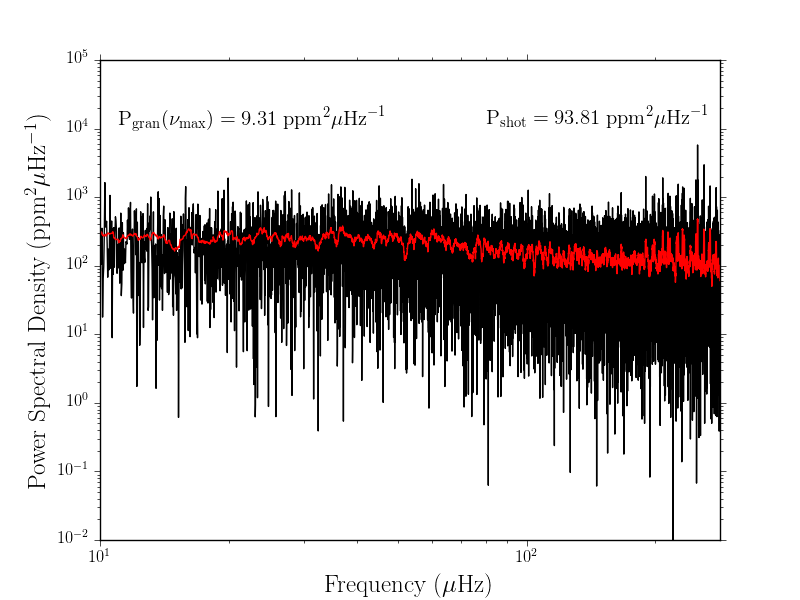}}
  \caption{\small Power spectra of Kepler-56 as seen both by \textit{Kepler} (left) and \textit{TESS} (right) based on 1 year of observations. A smoothed version of the power spectra is shown in red. Granulation power (at $\nu_{\rm max}$) and shot noise levels are indicated.}\label{fig:Kepler-56}
\end{figure*}

In Figure \ref{fig:Kepler-56} we provide a sneak peek at a power spectrum of a low-luminosity red-giant star as it would be observed by \textit{TESS} at the 30-min cadence. The two panels in this figure show the power spectra of Kepler-56 as observed both by \textit{Kepler} (in its long-cadence mode) and \textit{TESS} based on 1 year of observations. As can be seen, oscillations are easily identified in either power spectrum. Notice, however, that while the granulation power and shot noise levels are commensurate for the \textit{Kepler} power spectrum, the shot noise level is one order of magnitude higher than the granulation power in the case of the \textit{TESS} power spectrum. This is mostly due to the smaller (by 2 orders of magnitude) effective collecting area of the individual \textit{TESS} cameras.

\section{Summary and discussion}
\label{sec:summary}
In \cite{Campante16} we developed a simple test to estimate the detectability of solar-like oscillations in \textit{TESS} photometry of any given star. We applied this detection test here along a sequence of stellar evolutionary tracks in order to predict the detectability of solar-like oscillations across the H--R diagram (Section \ref{sec:detectability}). Importantly, we note that bright subgiants are attractive targets for the 2-min-cadence slots reserved for asteroseismology, even though they are far from optimal for transiting planet detection. We should be able to use parallaxes from the ongoing \textit{Gaia} mission \cite{Gaia} to deliberately target these bright subgiants.

\begin{acknowledgement}
The authors would like to thank the organizers of the ``Seismology of the Sun and the Distant Stars 2016'' conference, a joint TASC2/KASC9 workshop -- SPACEINN \& HELAS8 conference, where the author presented an invited talk on the contents of these proceedings. The author acknowledges the support of the UK Science and Technology Facilities Council (STFC). Funding for the Stellar Astrophysics Centre is provided by The Danish National Research Foundation (Grant DNRF106).
\end{acknowledgement}

\bibliography{biblio}
%
%
%
%

\end{document}